# Computational investigation of formation enthalpies and phase stability for rare earth oxyphosphates

Edric X. Wang[a,b], Ligen Wang[a], Qi-Jun Hong[a,]*
[a] School for Engineering of Transport, Energy and Matter, Arizona State University, Tempe, AZ 85287, USA
[b] Department of Materials Science and Engineering, University of Illinois at Urbana-Champaign, Urbana, IL 61801, USA

## Abstract

Rare earth phosphates have garnered significant interest due to their versatile properties, including high chemical stability, thermal resistance, luminescence, and the ability to adopt various crystalline structures. Density functional theory (DFT)-based ab initio methods have become essential tools for complementing experimental studies. In this paper, we performed DFT calculations on rare earth (RE; here considered as lanthanides + Y) oxyphosphates to examine their formation enthalpies and phase stability. The calculations were conducted using the GGA-PBE and r2SCAN exchange-correlation functionals. Our results indicate that both functionals predict similar phase stabilities for $REPO_4$ and $RE_3PO_7$. However, the r2SCAN functional provides significantly more accurate formation enthalpies for the monazite and xenotime $REPO_4$, aligning closely with experimental data. Furthermore, the inclusion of lattice vibrational entropy enhances the free energy predictions, leading to improved agreement with experimental observations on phase stability.

# 1. Introduction

Rare earth orthophosphates, such as monazite and xenotime, are among the most refractory minerals known, with extensive applications ranging from laser hosts to actinide immobilization matrices. In contrast, their synthetic counterparts, rare earth oxyphosphates (REOPs), have been largely overlooked since their discovery nearly half a century ago [1], despite their potential significance. REOPs are characterized by the general formula $(RE_2O_3)xRE(PO_4)$, with RE/P ratios of 7:3, 3:1, 4:1, and 5:1. Rare earth oxyphosphates have been successfully synthesized using methods such as co-precipitation followed by high-temperature annealing, and laser melting of $REPO_4$ [2]. Both experimental and computational approaches are being used to explore the structures and thermodynamic and physical properties of REOPs [1-11].

Rare earth oxyphosphates are characterized by their unique structural, optical, and thermal properties, making them valuable in a wide range of scientific and industrial applications. The interest in rare earth oxyphosphates stems from their high chemical stability, thermal resistance, luminescent capabilities, and ability to form a variety of crystalline structures, which can be tailored for specific uses in catalysis, luminescence, environmental science, material science, and biomedicine [3-9].

Computational materials science has made significant progress over the past few decades, largely driven by advancements in computational power and the development of ab initio methods, such as density functional theory (DFT) [12-14]. These methods allow researchers to simulate the atomic-scale behavior of materials, providing insights into their chemical and physical properties. This computational approach complements experimental techniques, enabling the exploration of atomic mechanisms responsible for material performance, such as stability, phase transitions, and reaction pathways. By modeling atomic interactions with increasing accuracy, computational methods have substantially expanded our understanding of materials, aiding in the design of more efficient and advanced materials across various applications [14].

In this study, we conducted a computational analysis of the formation enthalpies and phase stability of rare earth phosphates. Using density functional theory with the GGA-PBE [15] and r2SCAN [16] exchange-correlation functionals, we found that both functionals predicted similar phase stabilities for $REPO_4$ and $RE_3PO_7$. However, the r2SCAN functional yielded significantly more accurate formation enthalpies for monazite and xenotime $REPO_4$, closely matching experimental results. Thermal effects, such as thermal expansion and lattice vibrational entropy, were also calculated to determine free energy, which improved phase stability predictions and enhanced agreement with experimental observations.

## 2. Computational methods

The calculations are performed within the framework of DFT using the Vienna Ab-initio Simulation Package (VASP) [17,18]. Projector augmented wave (PAW) pseudopotentials [19] are employed to describe the interactions between ions and valence electrons. For the lanthanides Ce–Lu, pseudopotentials with a specific valence state of 3 are used, where the number of f-electrons frozen in the core is equal to the total number of valence electrons minus the formal valency [20]. For example, according to the periodic table, Sm has 8 valence electrons in total—6 f-electrons and 2 s-electrons. In most compounds, Sm typically exhibits a valency of 3, meaning that when the Sm_3 pseudopotential is generated, 5 f-electrons are treated as core electrons [20]. In density functional theory calculations, the generalized gradient approximation (GGA) for the exchange-correlation functional is widely used, offering a balance between computational efficiency and accuracy. Recent large-scale benchmarking of approximately 6,000 solid materials has shown that the r2SCAN exchange-correlation functional delivers both high numerical efficiency and accuracy, making it a preferred choice for many applications [21,22]. Specifically, r2SCAN provides more accurate formation enthalpies than GGA, the SCAN [23] and PBEsol [24] functionals for materials with both strong and weak bonding. In this work, we used the r2SCAN functional and compared its results with those obtained using the GGA-PBE functional. The plane wave basis set had an energy cutoff of 520 eV. The Brillouin zone was sampled with k-point meshes [25] that were similar in structure but denser than those used in the Materials Project database [11]. The total energy convergence was carefully tested with respect to the density of the k-point grid. Supercell parameters and atomic positions were optimized until the total energy converged within $10^{-5}$ eV, and atomic forces were minimized to below 0.02 eV/Å.

The enthalpies of formation of monazite and xenotime $REPO_4$ are calculated assuming the reaction: $\frac{1}{2}RE_2O_3 + \frac{1}{2}P_2P_5 \rightarrow REPO_4$. The formation enthalpy for $REPO_4$ with respect to $RE_2O_3$ and $P_2O_5$ oxides is defined as

$$\Delta E_f^{ox} = \mathrm{E_{tot}}(\mathrm{REPO_4}) - \frac{1}{2}\mathrm{E_{tot}}(\mathrm{RE_2O_3}) - \frac{1}{2}\mathrm{E_{tot}}(P_2\mathrm{O_5}), \quad (1)$$

where $\mathrm{E_{tot}}(\mathrm{REPO_4})$, $\mathrm{E_{tot}}(\mathrm{RE_2O_3})$, and $\mathrm{E_{tot}}(P_2\mathrm{O_5})$ are the total energies for the bulk solid phases.

Stability with respect to phase separation can be assessed using the convex hull formalism [26]. To assess the (in)stability of $RE_3PO_7$ phase, the reaction energy for forming $RE_3PO_7$ from the linear combination of neighbors ($RE_2O_3$ and $REPO_4$) in composition space is defined as:

$$\Delta \mathrm{E_f} = \mathrm{E_{tot}}(\mathrm{RE_3PO_7}) - \mathrm{E_{tot}}(\mathrm{RE_2O_3}) - \mathrm{E_{tot}}(\mathrm{REPO_4}), \quad (2)$$

where $\mathrm{E_{tot}}(\mathrm{RE_3PO_7})$, $\mathrm{E_{tot}}(\mathrm{RE_2O_3})$, and $\mathrm{E_{tot}}(\mathrm{REPO_4})$ are the total energies for $RE_3PO_7$, $RE_2O_3$, and $REPO_4$, respectively. For an unstable compound that lies above the convex hull, $\Delta \mathrm{E_f} > 0$, whereas for a stable compound that lies on or below the convex hull, $\Delta \mathrm{E_f} \leq 0$.

# 3. Results and discussions

## 3.1 Structure and energetics of rare earth sesquioxides

Rare earth sesquioxides exhibit various polymorphs across different temperature ranges, including the cubic (C), monoclinic (B), and hexagonal (A) phases at low temperatures, as well as the high-temperature hexagonal (H) and cubic (X) phases [27,28]. Table 1 presents the calculated energies for RE oxides in the low-temperature A, B, and C structures. For all sesquioxides, the C phase has the lowest total energy, establishing it as the ground state based on GGA-PBE calculations. Using the r2SCAN exchange-correlation functional, the C phase remains stable for all $RE_2O_3$ compounds except $La_2O_3$ and $Ce_2O_3$. As shown in Table 1, the A and C phases of $La_2O_3$, $Ce_2O_3$, and $Pr_2O_3$ are nearly isoenergetic. These predictions using the r2SCAN exchange-correlation functional align more closely with experimental observations, as the light rare earth oxides—$La_2O_3$, $Ce_2O_3$, $Pr_2O_3$, and $Nd_2O_3$—are known to be most stable in the hexagonal A-type structure [27,28]. Table 2 and Fig. 1 present the optimized structural parameters and molar volumes for the C-type $RE_2O_3$ oxides. For comparison, experimental values obtained from X-ray measurements and previously calculated results [29] are also included. The lattice constants for all RE oxides, except $Ce_2O_3$, are accurately predicted by both the GGA-PBE and r2SCAN functionals, with errors less than 1% relative to experimental values. For $Ce_2O_3$, the prediction errors are approximately 2.7% (GGA-PBE) and 2.5% (r2SCAN), respectively. The discrepancy between the calculated and experimental lattice constants for $Ce_2O_3$ may result from the presence of the $Ce^{4+}$-related phase and/or intermediate phases between $Ce_2O_3$ and $CeO_2$ in the experimental samples [30,31]. The r2SCAN functional systematically predicts smaller lattice constants than the GGA-PBE functional. Compared to experimental data, r2SCAN provides better accuracy for material systems with larger rare earth elements (Y and those earlier than Eu) but less accuracy for oxides with smaller rare earth metals (Gd through Lu). Our GGA-PBE results for both phase stability and lattice constants agree well with previously reported GGA-PBE calculations [29]. Together, these structural parameters and the formation enthalpy results presented below suggest that the r2SCAN exchange-correlation functional offers a more accurate description of rare earth oxides and oxyphosphates than the GGA-PBE functional.

**Table 1.** Total energies for $RE_2O_3$ (eV/formula unit) calculated by the GGA-PBE and r2SCAN functionals.

| | GGA-PBE | | | r2SCAN | | |
|---|---|---|---|---|---|---|
| | A-type | B-type | C-type | A-type | B-type | C-type |
| $La_2O_3$ | -41.825 | -41.799 | -41.948 | -96.057 | -96.006 | -96.051 |
| $Ce_2O_3$ | -40.586 | -40.572 | -40.697 | -94.517 | -94.481 | -94.505 |
| $Pr_2O_3$ | -40.827 | -40.824 | -40.955 | -94.718 | -94.693 | -94.725 |
| $Nd_2O_3$ | -41.004 | -41.013 | -41.154 | -95.025 | -95.011 | -95.054 |
| $Sm_2O_3$ | -41.217 | -41.253 | -41.415 | -96.041 | -96.056 | -96.125 |
| $Eu_2O_3$ | -41.368 | -41.417 | -41.590 | -97.095 | -97.162 | -97.258 |
| $Gd_2O_3$ | -41.522 | -41.587 | -41.775 | -98.318 | -98.370 | -98.476 |
| $Tb_2O_3$ | -41.589 | -41.670 | -41.872 | -99.679 | -99.748 | -99.873 |
| $Dy_2O_3$ | -41.636 | -41.734 | -41.949 | -101.286 | -101.374 | -101.517 |
| $Y_2O_3$ | -45.411 | -45.529 | -45.771 | -77.757 | -77.868 | -78.042 |
| $Ho_2O_3$ | -41.677 | -41.790 | -42.019 | -103.161 | -103.270 | -103.431 |
| $Er_2O_3$ | -41.720 | -41.851 | -42.093 | -105.356 | -105.482 | -105.655 |
| $Tm_2O_3$ | -41.796 | -41.944 | -42.199 | -107.927 | -108.073 | -108.269 |
| $Yb_2O_3$ | -41.829 | -41.995 | -42.267 | -110.841 | -111.009 | -111.222 |
| $Lu_2O_3$ | -41.859 | -42.041 | -42.326 | -113.951 | -114.139 | -114.373 |

**Table 2.** Structural parameters for cubic C-type $RE_2O_3$ (volume in Å$^3$/formula unit). The measured structural parameters taken from Ref. [29] were gathered from various experimental studies.

| | Calculated (this work) | | | | Calculated (Ref. [29]) | | Measured | |
|---|---|---|---|---|---|---|---|---|
| | GGA-PBE | | r2SCAN | | GGA-PBE | | | |
| | *a* (Å) | V (Å$^3$) | *a* (Å) | V (Å$^3$) | *a* (Å) | V (Å$^3$) | *a* (Å) | V (Å$^3$) |
| $La_2O_3$ | 11.397 | 92.54 | 11.364 | 91.73 | 11.387 | 92.29 | - | - |
| $Ce_2O_3$ | 11.409 | 92.82 | 11.377 | 92.03 | 11.414 | 92.94 | 11.111 | 85.73 |
| $Pr_2O_3$ | 11.278 | 89.66 | 11.248 | 88.94 | 11.290 | 89.94 | - | - |
| $Nd_2O_3$ | 11.173 | 87.19 | 11.130 | 86.16 | 11.178 | 87.30 | - | - |
| $Sm_2O_3$ | 10.989 | 82.93 | 10.943 | 81.91 | 10.998 | 83.14 | 10.930 | 81.61 |
| $Eu_2O_3$ | 10.899 | 80.91 | 10.844 | 79.70 | - | - | - | - |
| $Gd_2O_3$ | 10.809 | 78.92 | 10.747 | 77.58 | 10.819 | 79.16 | 10.790 | 78.51 |
| $Tb_2O_3$ | 10.736 | 77.34 | 10.669 | 75.91 | 10.744 | 77.50 | 10.729 | 77.10 |
| $Dy_2O_3$ | 10.665 | 75.82 | 10.598 | 74.39 | 10.675 | 76.02 | 10.670 | 75.92 |
| $Y_2O_3$ | 10.655 | 75.60 | 10.611 | 74.67 | 10.701 | 76.58 | 10.596 | 74.36 |
| $Ho_2O_3$ | 10.602 | 74.49 | 10.527 | 72.92 | 10.609 | 74.63 | 10.580 | 74.02 |
| $Er_2O_3$ | 10.544 | 73.27 | 10.462 | 71.56 | 10.544 | 73.26 | 10.548 | 73.35 |
| $Tm_2O_3$ | 10.474 | 71.81 | 10.391 | 70.12 | 10.472 | 71.77 | 10.480 | 71.94 |
| $Yb_2O_3$ | 10.414 | 70.59 | 10.332 | 68.94 | - | - | - | - |
| $Lu_2O_3$ | 10.354 | 69.38 | 10.267 | 67.64 | - | - | - | - |

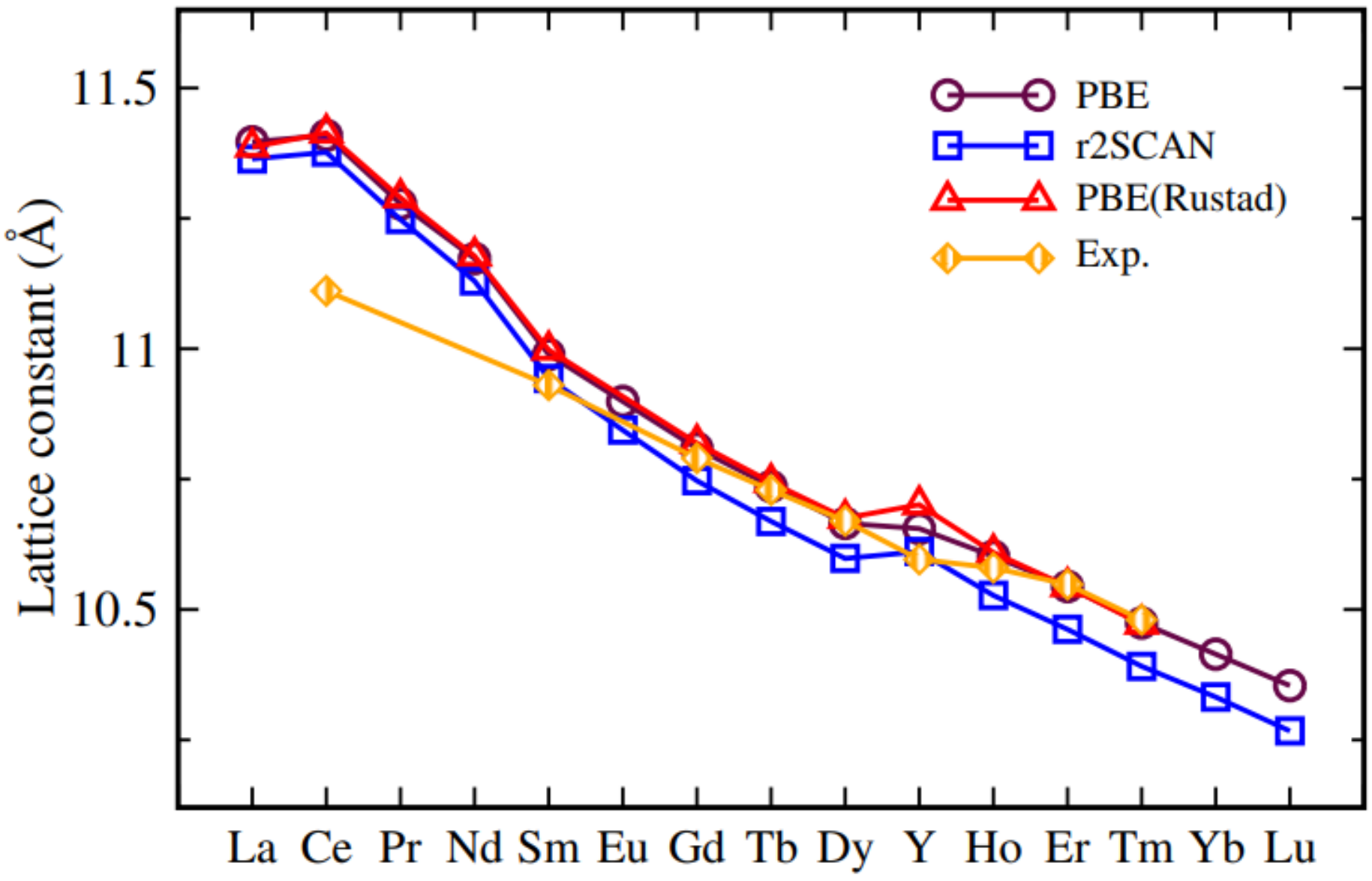

**Fig. 1.** Calculated lattice constant results for the C-type $RE_2O_3$ oxides, together with the experimental and previously reported theoretical values (Ref. [29]).

### 3.2 Phase stability and formation enthalpies of $REPO_4$

Rare earth phosphates, $REPO_4$, naturally occur in the monazite and xenotime phases. The formation enthalpies of rare earth orthophosphates from their oxides have been studied both experimentally [10] and computationally [29,32-34]. Previous studies identified two key issues when using density functional theory (DFT) to calculate the energetics of rare earth orthophosphates. First, DFT calculations inaccurately predict the xenotime phase to be more stable than the monazite phase for $EuPO_4$ and $SmPO_4$ [29]. For $NdPO_4$, prior DFT calculations with the GGA-PBE functional showed the monazite and xenotime structures to be nearly isoenergetic, with the monazite structure having slightly lower energy [29]. The second issue is that the calculated formation enthalpies are consistently less exothermic by approximately 40 kJ/mol compared to measured values [29,32-34]. This discrepancy is attributed to challenges in accurately describing phosphorus pentoxide and P–O bonds within DFT [32,34]. Here, formation enthalpy calculations for rare earth orthophosphates were performed using both the PBE and r2SCAN functionals. The r2SCAN functional has been shown to provide more accurate formation enthalpies across various materials, including oxides, rare earths, transition metal intermetallics and compounds, as well as materials with strong and weak bonding [21,22]. Thus, it is expected to deliver improved performance for rare earth orthophosphates. One advantage of theoretical calculations is the ability to investigate structures that have not been observed experimentally, such as xenotime $LaPO_4$ and monazite $LuPO_4$. Our calculations cover all monazite and xenotime $REPO_4$ structures, including those not experimentally observed. Formation enthalpies are calculated based on the reaction in Eq. (1).

Table 3 presents the total energies ($E^M$ and $E^X$) and energy differences ($\Delta E^{M-X}$) between the monazite and xenotime structures, along with previously reported results [29]. Subtle differences in total energies compared to prior calculations can be attributed to variations in computational parameters, despite both studies using the GGA-PBE exchange-correlation functional. The $\Delta E^{M-X}$ values from both GGA-PBE calculations show excellent agreement. However, the $\Delta E^{M-X}$ values obtained using the r2SCAN exchange-correlation functional differ notably from the GGA-PBE values, showing a stronger preference for the monazite structure. Importantly, both GGA-PBE and r2SCAN functionals predict the same phase stability trends: the monazite phase is more stable for La-, Ce-, Pr-, and Nd-orthophosphates, whereas the xenotime phase is more stable for all other orthophosphates (Sm–Lu and Y), as shown in Table 3 and Fig. 2. This prediction, however, appears inconsistent with experimental observations [10,35]. As discussed below, the monazite structure for certain orthophosphates ($SmPO_4$, $EuPO_4$, and $GdPO_4$) with small positive $\Delta E^{M-X}$ values becomes stable at finite temperatures.

The calculated formation enthalpies are presented in Fig. 3 and Table 4, alongside experimental formation enthalpies ($\Delta H_f^{ox}$ at 298 K) [10] for comparison. We used metastable hexagonal $P_2O_5$ (R3c, 161) in our formation enthalpy calculations to match the reference phase used in the experimental formation enthalpy measurements by Ushakov et al. [10]. With the r2SCAN exchange-correlation functional, the h-$P_2O_5$ phase is 0.051 eV/atom higher in energy than the polymeric orthorhombic o′-$P_2O_5$ compound, consistent with the Materials Project (MP) database value [11] and greater than the energy difference of 0.022 eV/atom calculated using the GGA-PBE functional [29]. As shown in previous studies, GGA-PBE formation enthalpies tend to be less exothermic by approximately 40 kJ/mol relative to experimental $\Delta H_f^{ox}(298K)$ values, while r2SCAN results have a notably smaller average error (RMSE of 8.7 kJ/mol). The improvement in formation enthalpy accuracy with r2SCAN cannot be fully explained by the energy difference between h-$P_2O_5$ and o′-$P_2O_5$, indicating r2SCAN's overall effectiveness in providing more accurate formation enthalpies for rare earth orthophosphates. As found in previous DFT calculations [29,33,34], the largest errors for both GGA-PBE and r2SCAN functionals occur with monazite $GdPO_4$ and $TbPO_4$.

**Table 3.** Calculated energies (eV/formula unit) of $REPO_4$ in the monazite (M) and xenotime (X) structures

| | Calculated (this work) | | | | | | Calculated (Ref.[29]) | | |
|---|---|---|---|---|---|---|---|---|---|
| | GGA-PBE | | | r2SCAN | | | GGA-PBE | | |
| | $E^M$ | $E^X$ | $\Delta E^{M-X}$ | $E^M$ | $E^X$ | $\Delta E^{M-X}$ | $E^M$ | $E^X$ | $\Delta E^{M-X}$ |
| $LaPO_4$ | -48.373 | -48.250 | -0.123 | -82.615 | -82.373 | -0.242 | –48.337 | –48.215 | -0.122 |
| $CePO_4$ | -47.850 | -47.723 | -0.127 | -81.950 | -81.709 | -0.240 | –47.805 | –47.679 | -0.126 |
| $PrPO_4$ | -47.879 | -47.814 | -0.065 | -81.961 | -81.788 | -0.173 | –47.843 | –47.779 | -0.064 |
| $NdPO_4$ | -47.884 | -47.876 | -0.008 | -82.029 | -81.919 | -0.110 | –47.854 | –47.846 | -0.008 |
| $SmPO_4$ | -47.842 | -47.936 | 0.094 | -82.379 | -82.385 | 0.007 | –47.835 | –47.923 | 0.088 |
| $EuPO_4$ | -47.844 | -47.991 | 0.148 | -82.826 | -82.906 | 0.080 | - | - | - |
| $GdPO_4$ | -47.838 | -48.037 | 0.199 | -83.344 | -83.478 | 0.134 | –47.819 | –48.013 | 0.194 |
| $TbPO_4$ | -47.803 | -48.046 | 0.243 | -83.949 | -84.135 | 0.186 | –47.788 | –48.027 | 0.239 |
| $DyPO_4$ | -47.759 | -48.044 | 0.285 | -84.677 | -84.913 | 0.236 | - | –48.025 | - |
| $YPO_4$ | -49.622 | -49.915 | 0.293 | -72.922 | -73.157 | 0.235 | - | –49.760 | - |
| $HoPO_4$ | -47.713 | -48.039 | 0.326 | -85.540 | -85.824 | 0.284 | - | –48.022 | - |
| $ErPO_4$ | -47.675 | -48.040 | 0.365 | -86.571 | -86.904 | 0.332 | - | –48.031 | - |
| $TmPO_4$ | -47.652 | -48.060 | 0.408 | -87.788 | -88.174 | 0.386 | - | –47.991 | - |
| $YbPO_4$ | -47.604 | -48.048 | 0.444 | -89.190 | -89.616 | 0.426 | - | - | - |
| $LuPO_4$ | -47.544 | -48.031 | 0.486 | -90.652 | -91.128 | 0.476 | - | - | - |

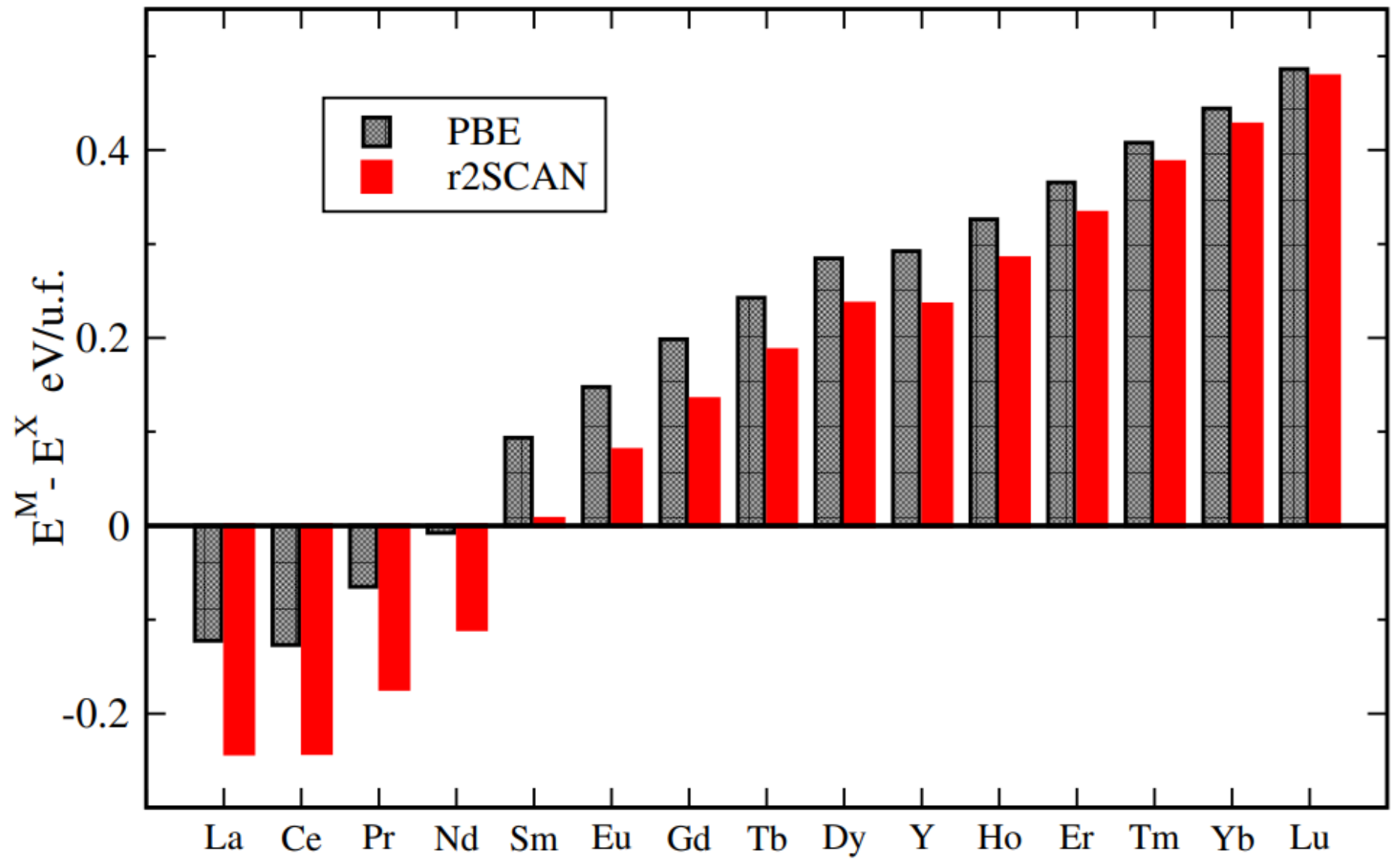

**Fig. 2.** Energy difference $\Delta E^{M-X}$ between the monazite and xenotime rare earth orthophosphates.

**Table 4.** Formation enthalpies $\Delta E_f^{ox}$(kJ/mol) at 0K with respect to oxides for $REPO_4$ in the monazite (M) and xenotime (X) structures

| | Measured $\Delta H_f^{ox}(298K)$ | Calculated $\Delta E_f^{ox}(0K)$ | | $\Delta E_f^{ox}(0K)$-$\Delta H_f^{ox}(298K)$ | |
|---|---|---|---|---|---|
| | | GGA-PBE | r2SCAN | GGA-PBE | R2SCAN |
| $LaPO_4$(m) | -321.4 | -279.4 | -321.7 | 42.0 | -0.3 |
| $LaPO_4$(x) | - | -267.6 | -298.6 | - | - |
| $CePO_4$(m) | -317.2 | -289.4 | -331.7 | 27.8 | -14.5 |
| $CePO_4$(x) | - | -277.1 | -309.1 | - | - |
| $PrPO_4$(m) | -312.2 | -279.7 | -322.8 | 32.5 | -10.6 |
| $PrPO_4$(x) | - | -273.5 | -306.1 | - | - |
| $NdPO_4$(m) | -312.0 | -270.6 | -313.4 | 41.4 | -1.4 |
| $NdPO_4$(x) | - | -269.9 | -302.8 | - | - |
| $SmPO_4$(m) | -301.8 | -254.0 | -295.6 | 47.8 | 6.2 |
| $SmPO_4$(x) | - | -263.0 | -296.2 | - | - |
| $EuPO_4$(m) | -286.8 | -245.7 | -284.1 | 41.1 | 2.7 |
| $EuPO_4$(x) | | -259.9 | -291.8 | - | - |
| $GdPO_4$(m) | -296.2 | -236.2 | -275.2 | 60.0 | 21.0 |
| $GdPO_4$(x) | - | -255.4 | -288.2 | - | - |
| $TbPO_4$(m) | -283.5 | -228.1 | -266.2 | 55.4 | 17.3 |
| $TbPO_4$(x) | -286.1 | -251.6 | -284.2 | 34.5 | 1.9 |
| $DyPO_4$(m) | - | -220.2 | -257.2 | - | - |
| $DyPO_4$(x) | -283.9 | -247.7 | -279.9 | 36.2 | 4.0 |
| $YPO_4$(m) | - | -215.6 | -255.5 | - | - |
| $YPO_4$(x) | -282.6 | -243.8 | -278.1 | 38.8 | 4.5 |
| $HoPO_4$(m) | - | -212.4 | -248.1 | - | - |
| $HoPO_4$(x) | -278.8 | -243.9 | -275.5 | 34.9 | 3.3 |
| $ErPO_4$(m) | - | -205.1 | -240.3 | - | - |
| $ErPO_4$(x) | -275.6 | -240.4 | -272.4 | 35.2 | 3.2 |
| $TmPO_4$(m) | - | -197.8 | -231.6 | - | - |
| $TmPO_4$(x) | -268.0 | -237.1 | -268.9 | 30.9 | -0.9 |
| $YbPO_4$(m) | - | -189.9 | -224.4 | - | - |
| $YbPO_4$(x) | -269.6 | -232.7 | -265.5 | 36.9 | 4.1 |
| $LuPO_4$(m) | - | -181.3 | -213.5 | - | - |
| $LuPO_4$(x) | -263.9 | -228.2 | -259.4 | 35.7 | 4.5 |
| RMSE | | | | 40.3 | 8.7 |

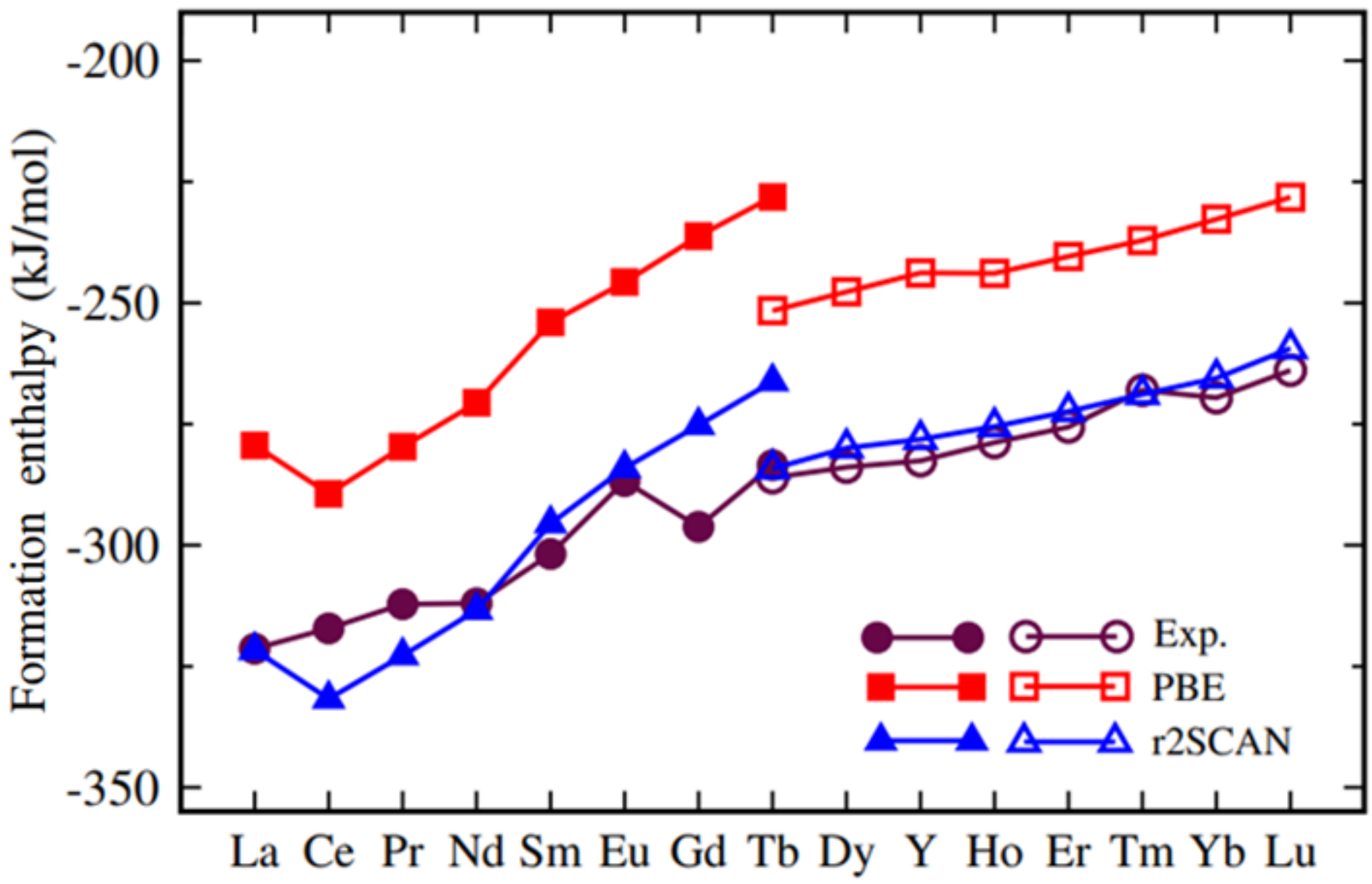

**Fig. 3.** The formation enthalpies of the monazite (filled symbols) and xenotime (open symbols) rare earth orthophosphates.

### 3.3 Formation of $RE_3PO_7$ from $RE_2O_3$ and $REPO_4$

Rare earth oxyphosphates have garnered significant research interest due to their promising functional applications in various fields, such as nuclear waste storage, phosphors, and solid-state lasers [6,7,9,36]. Oxyphosphates like $RE_3PO_7$ are known for their luminescent properties, particularly when doped with specific rare earth elements like Eu or Tb [9]. This makes them useful in various optical applications, such as LED lighting or display technologies. Additionally, their thermal and chemical stability has led to exploration in catalysis and high-temperature reactions [3]. These compounds feature a network of rare earth cations and phosphate anions, resulting in a structure that can host a variety of functional properties. To date, only a few oxyphosphates ($Nd_3PO_7$ and $Eu_3PO_7$) have had their structures experimentally identified as monoclinic (C2/m) using single-crystal XRD [4,5]. $Y_3PO_7$ was found to be isostructural with $Gd_3PO_7$, as reported by Tuan et al. [9]. From the MP database [11], four oxyphosphates ($La_3PO_7$, $Nd_3PO_7$, $Sm_3PO_7$, and $Tb_3PO_7$) are provided with a monoclinic (Cm) symmetry. In our calculations, we used the structures from the MP database and re-optimized the structures for all RE (lanthanides + Y) oxyphosphates. We found that the optimized structures for $La_3PO_7$, $Ce_3PO_7$, $Pr_3PO_7$, and $Nd_3PO_7$ retain the monoclinic Cm symmetry, while all other RE oxyphosphates exhibit monoclinic C2/m symmetry. The calculated formation enthalpies, based on the reaction in Eq. (2), are presented in Table 5 and Fig. 4. Fig. 4(b) shows the formation enthalpies at 0 K, calculated using the GGA-PBE and r2SCAN exchange-correlation functionals. For all $RE_3PO_7$ compounds from La through Dy, the formation enthalpies ($\Delta E_f$) are negative, indicating stability against decomposition into oxide and $REPO_4$, while other oxyphosphates tend to decompose. Thermal effects on phase stability are not included here and will be discussed below.

**Table 5.** Calculated total energies and formation enthalpies (in eV/formula unit) for $RE_3PO_7$. $\Delta E_f^M$ and $\Delta E_f^X$ are the formation enthalpy defined in Eq. (2) for the monazite and xenotime $REPO_4$, respectively. The value in bold font represents the formation enthalpy for the oxyphosphate $RE_3PO_7$, as the calculation uses the most stable $REPO_4$ phase (also see the filled circles in Fig. 4(a)).

| | GGA-PBE | | | r2SCAN | | |
|---|---|---|---|---|---|---|
| | $E_{tot}$ | $\Delta E_f^M$ | $\Delta E_f^X$ | $E_{tot}$ | $\Delta E_f^M$ | $\Delta E_f^X$ |
| $La_3PO_7$ | -90.656 | **-0.336** | -0.459 | -178.958 | **-0.286** | -0.528 |
| $Ce_3PO_7$ | -88.876 | **-0.329** | -0.459 | -176.725 | **-0.261** | -0.502 |
| $Pr_3PO_7$ | -89.166 | **-0.332** | -0.400 | -176.955 | **-0.269** | -0.442 |
| $Nd_3PO_7$ | -89.366 | **-0.328** | -0.339 | -177.351 | **-0.268** | -0.378 |
| $Sm_3PO_7$ | -89.576 | -0.323 | **-0.226** | -178.765 | -0.261 | **-0.254** |
| $Eu_3PO_7$ | -89.750 | -0.316 | **-0.165** | -180.328 | -0.243 | **-0.164** |
| $Gd_3PO_7$ | -89.922 | -0.311 | **-0.110** | -182.070 | -0.250 | **-0.116** |
| $Tb_3PO_7$ | -89.975 | -0.305 | **-0.057** | -184.066 | -0.244 | **-0.058** |
| $Dy_3PO_7$ | -89.998 | -0.294 | **-0.005** | -186.431 | -0.237 | **-0.001** |
| $Y_3PO_7$ | -95.652 | -0.263 | **0.033** | -151.144 | -0.181 | **0.055** |
| $Ho_3PO_7$ | -90.012 | -0.282 | **0.046** | -189.199 | -0.228 | **0.057** |
| $Er_3PO_7$ | -90.036 | -0.270 | **0.097** | -192.447 | -0.220 | **0.112** |
| $Tm_3PO_7$ | -90.109 | -0.258 | **0.149** | -196.263 | -0.206 | **0.180** |
| $Yb_3PO_7$ | -90.116 | -0.245 | **0.199** | -200.608 | -0.196 | **0.230** |
| $Lu_3PO_7$ | -90.110 | -0.238 | **0.246** | -205.210 | -0.185 | **0.291** |

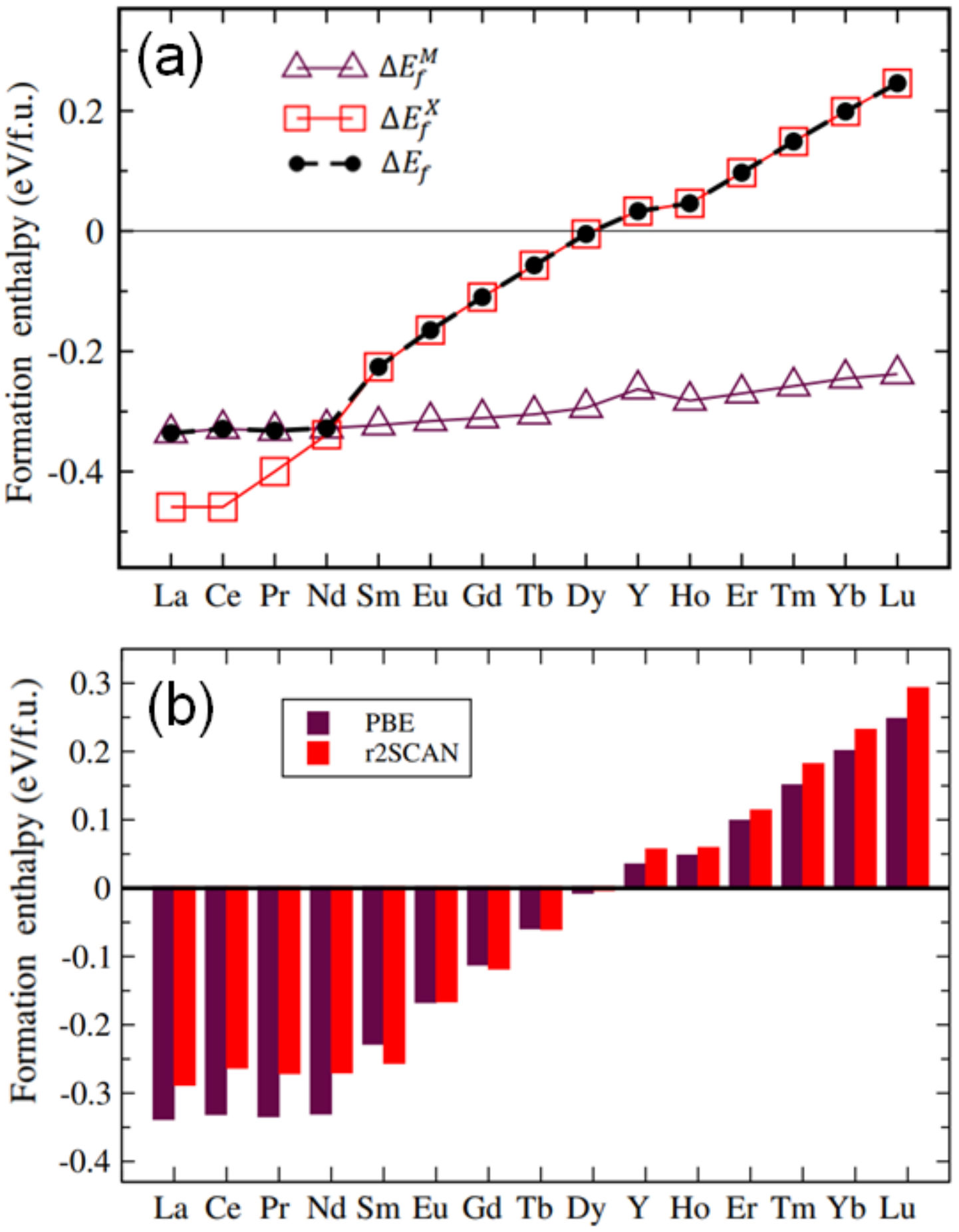

**Fig. 4.** (a) Illustrating the calculated formation enthalpy $\Delta E_f$ (with respect to the most stable

$REPO_4$ phase and the C-type oxide) for $RE_3PO_7$ obtained from the formation enthalpies $\Delta E_f^M$ and $\Delta E_f^X$. The results are presented for the case of the GGA-PBE exchange-correlation functional. (b) Calculated formation enthalpies $\Delta E_f$ using both GGA-PBE and r2SCAN exchange-correlation functionals.

### 3.4 Vibrational entropy contribution to phase stability

Lattice vibrations are known to have a significant impact on phase stability [37]. The difference in vibrational entropy between phases arises from factors related to atomic and bonding structures, including bond stiffness, volume changes, and atomic size mismatches. The ATAT software [38], developed by Axel van de Walle, provides the *fitfc* method for highly accurate calculations of vibrational free energies. This method fits a Born–von Kármán spring model to the reaction forces generated by imposed atomic displacements in a supercell calculation to determine vibrational properties. The VASP code needs to be invoked to calculate reaction forces for each perturbation. The r2SCAN exchange-correlation functional was employed in the VASP calculations. The *fitfc* calculations were performed for several orthophosphates to investigate the stability of the monazite structure relative to the xenotime structure, as there is a discrepancy between phase stabilities predicted from formation enthalpy calculations at 0 K (Table 3 and Fig. 2) and experimental observations. Figure 5 shows the free energy differences between the monazite and xenotime structures for Nd-, Sm-, Eu-, Gd-, Tb-, Dy-, Ho-, and Y-orthophosphates. It can be seen in Fig. 5 that the monazite phase is preferred for RE orthophosphates up to Sm, while from Dy to Lu (including Y), the xenotime phase is stable. For Eu-, Gd-, and Tb-orthophosphates, both the monazite and xenotime phases may be stabilized in different temperature regions. Therefore, after accounting for the vibrational entropy contribution to free energy, our theoretical predictions of phase stability align more closely with experimental observations.

Formation enthalpy calculations at 0 K have predicted that from Ho to Lu (including Y), the oxyphosphates $RE_3PO_7$ are unstable and decompose into $REPO_4$ and $RE_2O_3$, as shown in reaction Eq. (2). The vibrational entropy contributions at finite temperatures have been calculated by the *fitfc* method for the $RE_3PO_7$ oxyphosphates with near zero or slightly positive formation enthalpies at 0 K and Fig. 6 shows the temperature dependence of formation free energy [$\Delta G_f$, calculated by substituting total energies with free energies in Eq. (2)] for those $RE_3PO_7$ oxyphosphates. After accounting for the vibrational entropy contribution, $Y_3PO_7$, $Ho_3PO_7$, and $Er_3PO_7$ can be stabilized at finite temperatures, whereas Tm and later RE oxyphosphates remain unstable below 2200 K. These results, including finite temperature effects, agree well with the experimental observations for $RE_3PO_7$ phase stability [39].

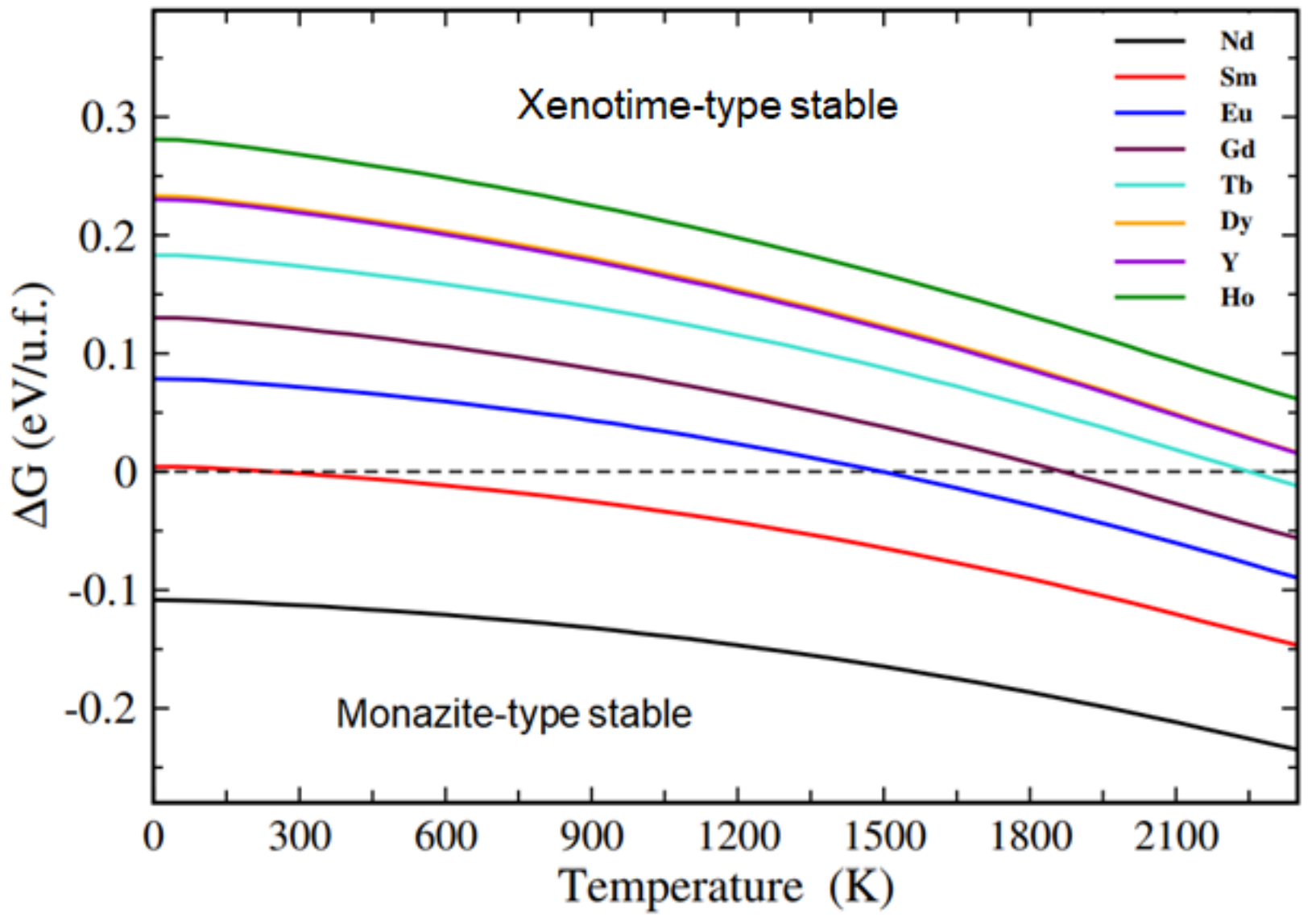

**Fig. 5.** Free energy difference between the monazite and xenotime phases $REPO_4$.

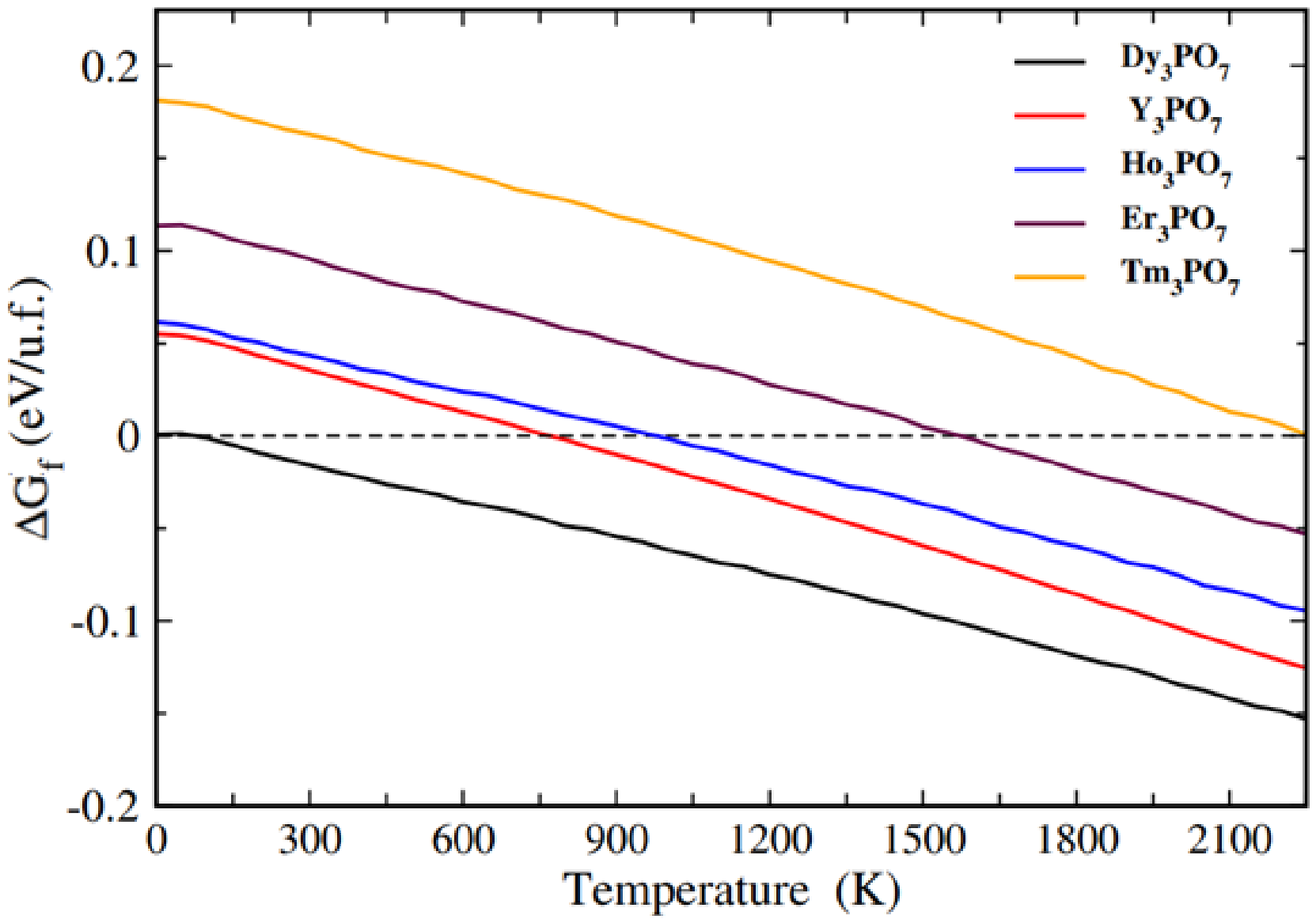

**Fig. 6.** The dependence of the formation free energy on temperature for oxyphosphates $RE_3PO_7$.

## 5. Conclusions

In conclusion, this study provides a comprehensive computational analysis of the formation enthalpies and phase stability of rare earth phosphates. Both GGA-PBE and r2SCAN functionals predict similar phase stabilities for $REPO_4$ and $RE_3PO_7$, with r2SCAN delivering notably accurate formation enthalpies for monazite and xenotime $REPO_4$, closely aligning with experimental results. Including thermal effects, particularly lattice vibrational entropy, further refines free energy predictions, enhancing the agreement between theoretical models and experimental observations on phase stability. These findings underscore the effectiveness of DFT-based methods, particularly with the r2SCAN functional, in accurately characterizing rare earth phosphates' stability and formation properties.

**Acknowledgments**

This research was supported by US Department of Defense Army Research Office Award number W911NF-23-2-0145, with use of Research Computing at Arizona State University.